\documentclass[a4paper,11pt]{article}

\usepackage{jinstpub} 

\usepackage{placeins}
\usepackage{lscape}

\title{First FAMU observation of muon transfer from $\mu$p atoms to higher-Z elements}
\author[a]{E.~Mocchiutti,\note{Corresponding author.}}
\author[a]{, V.~Bonvicini}
\author[a]{, R.~Carbone}
\author[a,b]{, M.~Danailov}
\author[a]{, E.~Furlanetto}
\author[a,c,d]{, K.S.~Gadedjisso-Tossou}
\author[a,e]{, D.~Guffanti}
\author[a]{, C.~Pizzolotto}
\author[a]{, A.~Rachevski}
\author[a,c]{, L.~Stoychev}
\author[a]{, E.~Vallazza}
\author[a]{, G.~Zampa}
\author[c]{, J.~Niemela}
\author[f]{, K.~Ishida}

\author[g]{, A.~Adamczak}
\author[h,i]{, G.~Baccolo}
\author[h,j]{, R.~Benocci}
\author[h]{, R.~Bertoni}
\author[h]{, M.~Bonesini}
\author[h]{, F.~Chignoli}
\author[h,i]{, M.~Clemenza}
\author[h]{, A.~Curioni}
\author[h,j]{, V.~Maggi}
\author[h]{, R.~Mazza}
\author[h,i]{, M.~Moretti}
\author[h,i]{, M.~Nastasi}
\author[h]{, E.~Previtali}
\author[k]{, D.~Bakalov}
\author[k]{, P.~Danev}
\author[k]{, M.~Stoilov}
\author[l,m]{, G.~Baldazzi}
\author[l]{, G.~Campana}
\author[l]{, I.~D'Antone}
\author[l]{, M.~Furini}
\author[l,n]{, F.~Fuschino}
\author[l,n]{, C.~Labanti}
\author[l]{, A.~Margotti}
\author[l]{, S.~Meneghini}
\author[l,n]{, G.~Morgante}
\author[l,m]{, L.~P.~Rignanese}
\author[l]{, P.L.~Rossi}
\author[l]{, M.~Zuffa}
\author[o,p]{, T.~Cervi}
\author[o,p]{, A.~De~Bari}
\author[o,p]{, A.~Menegolli}
\author[p]{, C.~De~Vecchi}
\author[p]{, R.~Nard\`o}
\author[p]{, M.~Rossella}
\author[p,q]{, A.~Tomaselli}
\author[r,s]{, L.~Colace}
\author[r,t]{, M.~De~Vincenzi}
\author[r]{, A.~Iaciofano}
\author[r,u]{, F.~Somma}
\author[r]{, L.~Tortora}
\author[v]{, R.~Ramponi}
\author[a,f,w]{, and A.~Vacchi}

\affiliation[a]{National Institute for Nuclear Physics (INFN), Sezione di
Trieste, via A. Valerio 2, 34127 Trieste, Italy}
\affiliation[b]{Elettra-Sincrotrone Trieste S.C.p.A., SS14, Km 163.5, 34149
Basovizza, Trieste, Italy}
\affiliation[c]{The Abdus Salam International Centre for Theoretical Physics,
  Strada Costiera 11, Trieste, Italy}
\affiliation[d]{Laboratoire de Physique des Composants \`a Semi-conducteurs (LPCS),
D\'epartement de physique, Universit\'e de Lom\'e, 01 BP 1515 Lom\'e, Togo}
\affiliation[e]{Gran Sasso Science Institute - INFN-LNGS, via F. Crispi 7,
L'Aquila, Italy}
\affiliation[f]{RIKEN-RAL RIKEN Nishina Center for Accelerator-Based Science, 2-1, Hirosawa, Wako, Saitama 351-0198, Japan}
\affiliation[g]{Institute of Nuclear Physics, Polish Academy of Sciences,
Radzikowskiego 152, PL31342 Krak\'{o}w, Poland}
\affiliation[h]{National Institute for Nuclear Physics (INFN), Sezione di Milano
  Bicocca, piazza della Scienza 3, Milano, Italy}
\affiliation[i]{Universit\'a di Milano Bicocca, Dip. di Fisica ``G. Occhialini'', piazza della Scienza 3, Milano, Italy}
\affiliation[j]{Universit\'a di Milano Bicocca, Dip. di Scienze dell'Ambiente e della Terra, piazza della Scienza 1, Miano, Italy}
\affiliation[k]{Institute for Nuclear Research and Nuclear Energy, Bulgarian
Academy of Sciences, blvd. Tsarigradsko ch. 72, Sofia 1142, Bulgaria}
\affiliation[l]{National Institute for Nuclear Physics (INFN), Sezione di
  Bologna, viale Berti Pichat, 6/2, Bologna, Italy}
\affiliation[m]{Department of Physics and Astronomy, University of Bologna, viale Berti Pichat, 6/2, Bologna, Italy}
\affiliation[n]{INAF-IAFS Bologna, Area della Ricerca, via P. Gobetti 101, Bologna, Italy}
\affiliation[o]{Department of Physics, University of Pavia, via Bassi 6, Pavia,
Italy}
\affiliation[p]{National Institute for Nuclear Physics (INFN), Sezione di Pavia,
via Bassi 6, Pavia, Italy}
\affiliation[q]{Department of Electrical, Computer, and Biomedical Engineering,
University of Pavia, via Ferrata 5, Pavia, Italy}
\affiliation[r]{National Institute for Nuclear Physics (INFN), Sezione di Roma
Tre, via della Vasca Navale 84, Roma, Italy}
\affiliation[s]{Dipartimento di Ingegneria Universit\`a degli Studi Roma Tre, via V. Volterra, 62, Roma, Italy}
\affiliation[t]{Dipartimento di Matematica e Fisica, Universit\`a di Roma Tre,
via della Vasca Navale 84, Roma, Italy}
\affiliation[u]{Dipartimento di Scienze, Universit\`a di Roma Tre, viale G.
Marconi 446, Roma, Italy}
\affiliation[v]{IFN-CNR, Department of Physics - Politecnico di Milano and
National Institute for Nuclear Physics (INFN), Sezione di Milano Politecnico,
piazza Leonardo da Vinci 32, 20133 Milano, Italy}
\affiliation[w]{Mathematics and Informatics Department, Udine University, via
  delle Scienze 206, Udine, Italy}

\emailAdd{Emiliano.Mocchiutti@ts.infn.it}

\abstract{
The FAMU experiment aims to accurately measure the hyperfine splitting
of the ground state of the muonic hydrogen atom. A measurement of the
transfer rate of muons from hydrogen to heavier gases is necessary for
this purpose. In June 2014, within a preliminary experiment, a
pressurized gas-target was exposed to the pulsed low-energy muon beam
at the RIKEN RAL muon facility (Rutherford Appleton Laboratory,
UK). The main goal of the test was the characterization of both the
noise induced by the pulsed beam and the X-ray detectors. The
apparatus, to some extent rudimental, has served admirably to this
task. Technical results have been published that prove the validity of
the choices made and pave the way for the next steps. This paper
presents the results of physical relevance of measurements 
of the muon transfer rate to carbon dioxide, oxygen, and argon from
non-thermalized excited $\mu$p atoms. The
analysis methodology and the approach to the systematics errors are
useful for the subsequent study of the transfer rate as function of
the kinetic energy of the $\mu$p currently under way.  
}

\keywords{Muonic hydrogen; Tranfer rate; Oxygen}
\collaboration[c]{(FAMU collaboration)}

\begin{document}
\maketitle
\flushbottom

\section{The FAMU project}
\label{intro}
The final objective of the FAMU experiment is to determine the proton 
Zemach radius by measuring the hyperfine splitting of the $\mu$p ground 
state. In literature, the ``standard'' measurement of the Zemach
radius of the proton $R_p$ is achieved using ordinary hydrogen. A comparison with the value extracted from muonic hydrogen 
may reinforce or delimit the proton radius puzzle \cite{pohl10}.

The FAMU experimental method requires a detection system suited 
for time resolved X-ray spectroscopy \cite{adamczak16}. The characteristic X-rays from muonic atoms formed in 
different targets have been detected using HPGe detectors and five 
scintillating counters based on LaBr3(Ce) crystals, whose output was 
recorded for 5~$\mu$s using a 500 MHz digitizer to measure both energy and 
time spectrum of the detected events.  
 The expected characteristic muonic X-rays of various elements and the
 mean lifetimes were determined through a detailed pulse analysis.

In the proposed laser spectroscopy experiment, 
muonic hydrogen atoms are formed in a 
hydrogen gas target. In subsequent collisions with $H_2$ molecules, the
$\mu$p de-excite at thermalized $\mu$p in the $(1S)_{F=0}$ state. A laser tuned on 
the hyperfine splitting resonance induces singlet-to-triplet transitions; when the 
$\mu$p atoms in the $(1S)_{F=1}$ state are de-excited back to the singlet state 
and the transition energy is converted into additional kinetic energy of 
the $\mu$p system. Thus the $\mu$p atom gains about two-thirds of the hyperfine 
transition energy ($\approx 120$~meV). The energy dependence of the
muon transfer from muonic hydrogen to another higher-Z gas is
exploited to detect the transition which occurred in the $\mu$p. Although, in general, the 
muon-transfer rate at low energies $\Lambda_{pZ}$ is energy independent, this is 
not the case for a few gases. Oxygen exhibits a peak in the muon 
transfer rate $\Lambda_{pO}$ at epithermal
energy~\cite{werthmuller96}. Thus by adding small quantities of oxygen to hydrogen it is possible to
observe, by measuring the time distribution of the characteristic
X-rays from the muon-transfer events to the added gas, the number of
hyperfine transitions which take place. 
The setup of this first FAMU experimental test is described in details in \cite{adamczak16}:
it consisted in a beam hodoscope placed in front of a high pressure gas in an aluminium vessel surrounded by X-rays detectors. 
For these tests, four different targets were exposed to the muon beam: a pure 
graphite block and three gas mixtures (pure $H_2$, $H_2+2\%Ar$,
$H_2+4\%CO_2$, measured by weight). 
The hydrogen gas contained a natural
admixture of deuterium.
The aim was to study detector response 
in the environment of the muon beam at RIKEN RAL through the measurement 
of the muon transfer rate at room temperature. The  muonic atom characteristic X-rays  were detected using scintillating counters based on LaBr3(Ce) crystals  (energy resolution 
$\approx$3\% at 662 keV and decay time $\tau=16$~ns) read out by Hamamatsu R11265-200 PMTs and two HPGe detectors were used to obtain a benchmark spectrum. The waveforms were processed off-line to reconstruct the time and energy of each detected X-ray.

\section{Muonic atom formation  and muon transfer}
\label{sec:1}
At RAL muons are produced in bunches with a repetition rate of 50~Hz. Each
bunch is consisting of two spills separated by about 320~ns. Each spill can
be roughly described as a gaussian distribution with FWHM of 70~ns
and the incident muon momentum has a gaussian distribution with
$\sigma_E/E \approx 10\%$~\cite{RAL}.
The mean of the distribution can be tuned by beam users
and it was chosen to be 61~MeV/c for the 2014 FAMU data acquisition.

Exiting the kapton window of the beam pipe collimator at a rate of about $10^5$ particle per second, muons cross the FAMU beam hodoscope~\cite{bonesini17} and reach the gas in the aluminium vessel after crossing a 3~mm thick aluminium window.

Muons loose energy mostly by the ionization process.
The initial momentum was
chosen in order to maximize the muon stop in the gas of the
target. Once a muon is slowed down to zero, it is
captured by the closest atom into high orbital states, forming a
muonic atom. 

The muon atomic orbit equivalent to the electron $K$ orbit has a
principal quantum number
$n_\mu \approx (m_\mu/m_e)^{(1/2)} \approx 14$
and it is reached by the muon in femtoseconds from the
instant of its atomic capture \cite{mukhopadhyay}. The muon cascades down rapidly to the lowest
quantum state available, $1S$. In the case of light elements,
de-excitation starts with Auger process until the quantum number
reaches values from 3 to 6 at which radiative
transitions take over. During
radiative transitions X-rays are emitted at an energy corresponding to
the energy levels difference. Due to momentum conservation, muonic
hydrogen gains kinetic energy during this phase.

In pure hydrogen, once the muon is in the $1S$ orbit it either decays ($\mu^- \rightarrow e^-+ \overline \nu_e + \nu_\mu$) or is captured by
the nucleus ($\mu^- + p \rightarrow n + \nu_\mu$). In hydrogen the
capture to decay probability ratio is of
the order of $4\times 10^{-4}$, while this ratio becomes about one
around $Z=11$ \cite{mukhopadhyay}.

In presence of material impurities with mass number A$>$1 and atomic number Z$>$1,
the muonic hydrogen system can undergo the transfer reaction
$\mu^- p + ~^{A}\!Z \rightarrow$ $ ({\mu^-} ~^A\!Z)^* + p$, which results in the
formation of a new muonic atom $({\mu^-} ~^A\!Z)^*$ which then de-excites by
Auger and radiative processes. 

The muon transfer rate from hydrogen to higher-Z gases depends on the
density of the gas, on the amount of impurities in the hydrogen gas
and on the kinetic energy of the muonic hydrogen.

Once constructed, to avoid any contamination, the aluminium target was
submitted to cycles of washing and heating. During transportation the
target was filled with nitrogen at about 2 bar. For every gas change,
the target was evacuated at a vacuum level below $10^{-5}$~mbar.  This
procedure assured a contamination level smaller than one part per
million during the data taking. HpGe detectors confirmed the purity of
the gas inside the target.

\section{Data analysis}
\label{sec:2}
The measurement of the transfer rate as function of the muonic hydrogen
kinetic energy should be performed in stable thermalized condition. In
this state, at a given temperature of the target, the kinetic energy
of the $\mu$p follows the Maxwell-Boltzmann distribution.

Although the 2014 apparatus was devoted to study the beam conditions
and the signal to noise ratio attainable with the chosen detection system
(LaBr3(Ce) and HPGe detectors), a detailed analysis has been carried out
to measure the transfer rate from hydrogen to atoms with higher Z.

In this study the gas target was kept at a pressure of 38~bar at room
temperature. In these conditions the thermalization of the muonic
hydrogen, after formation, requires about 100
ns~\cite{bakalov15}. However at the  concentration  of $CO_2$ (4\% by
weight) used in the gas mixture the population of muonic hydrogen is
greatly suppressed by muon transfer to the $CO_2$ within times well
below 100 ns~\cite{bakalov15}. Hence the transfer arises in 
epithermal uncontrolled conditions at which the kinetic energy of
muonic hydrogen during the muon transfer can be assumed to be
distributed as a sum of two Maxwellians distributions, the first
related to the gas temperature and the second one corresponding to the
mean energy of 20 eV, (see \cite{werthmuller98} and references
therein). 

The 2014 data analysis showed a strong
X-ray emission from K-lines de-excitation of muonic oxygen and carbon
and at the same time no delayed oxygen or carbon lines emission after
about 10--20 ns from the arrival of the last of the muon beam
spills~\cite{adamczak16}.

 Since most of the muons are captured by the
hydrogen in the mixture -- as confirmed by a dedicated GEANT4
simulation~\cite{agostinelli} -- and the percentage of oxygen atoms is
order of per mille, the detected oxygen and carbon lines mostly come
from muon transfer from hydrogen to $CO_2$. 

Hence, a study of the time
evolution of the oxygen K-lines emission can be performed to measure
the transfer rate. Since the transfer process is faster than 100~ns, this
analysis concerns the X-rays from carbon and oxygen detected during
the arrival of the muon beam itself within the prompt X-ray background
emission from all the elements of the target, mostly aluminium. In this
situation  the  $\mu$p can be excited and have a very wide energy
distribution, so the level of epithermicity is undetermined.

The variation of the number of muonic hydrogen atoms $N_{\mu p}$ present in the target in the time interval $dt$ can be
expressed by:
\begin{equation}
  dN_{\mu p}(t) = S(t) dt - N_{\mu p}(t)\lambda_{dis} dt,
\label{eq:tevol}
\end{equation}
where $S(t)$ is the number of muonic hydrogen generated in the
time interval $dt$, and $\lambda_{dis}$ is the total disappearance
rate of the muonic hydrogen atoms:
\begin{equation}
\lambda_{dis} = \lambda_0+\phi\,(c_p\Lambda_{pp\mu}+  c_d\Lambda_{pd} +
c_{Z_1}\Lambda_{pZ_1}+c_{Z_2}\Lambda_{pZ_2}+\ldots).
\label{eq:ldis}
\end{equation}
Here $\lambda_0$ is the rate of disappearance of the muons
bounded to proton (that includes both muon decay and nuclear
capture), $\Lambda_{pp\mu}$ is the formation rate of the $pp\mu$ molecular ion in
collision of $\mu p$ with a hydrogen nucleus, $\Lambda_{pd}$ denotes the
muon transfer rate from $\mu$p to deuterium, 
 and $\Lambda_{pZ_i}$ are the muon transfer
 rates from $\mu p$ to the admixture nuclei of charge
 $Z_i$ ($i=1, 2, \ldots$ etc).
 The $pp\mu$ formation and muon transfer rates
 are all normalized to the liquid hydrogen number density (LHD)
 $N_0=4.25\times 10^{22}$~cm$^{-3}$, and $\phi$ is the target gas
 number density in LHD units.
 The atomic concentrations of hydrogen, deuterium, and
heavier-nuclei admixture in the gas target, denoted by
 $c_p$, $c_d$, $c_{Z_i}$,
are related to the number
densities of the latter, $N_p$, $N_d$, and $N_{Z_i}$, by:
\[ c_p=N_p/N_{tot} ,\: c_d=N_d/N_{tot}\]
\[ c_{Z_i}=N_{Z_i}/N_{tot} \:  (i=1,2,\ldots), \] 
\[N_{tot}=N_p+N_d+N_{Z_1}+N_{Z_2}+\ldots,\ \]
\[  \ c_p+c_d+\sum\,c_{Z_i}=1. \]
 The unknowns in Eq.~\ref{eq:tevol} and \ref{eq:ldis} are therefore $S(t)$ and
 $\Lambda_{pZ_i}$.

The key point of this analysis is that the time evolution of the emission of prompt X-rays from the aluminium of the target resembles the time shape of the beam pulses and can be used as a description of the formation rate of muonic hydrogen $S(t)$.

\begin{figure}[!htb]
\centering
\includegraphics[width=0.45\textwidth]{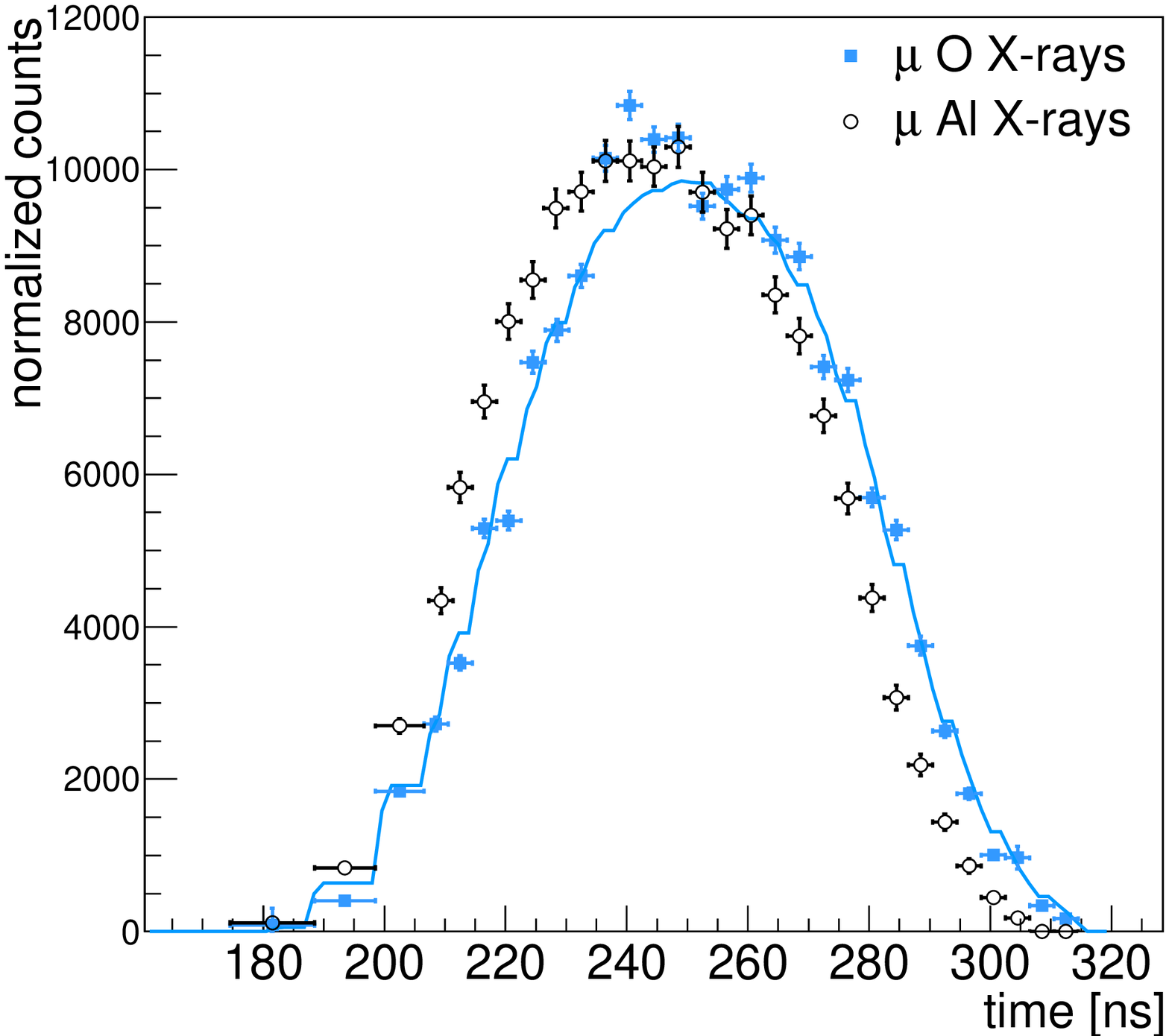}
\includegraphics[width=0.45\textwidth]{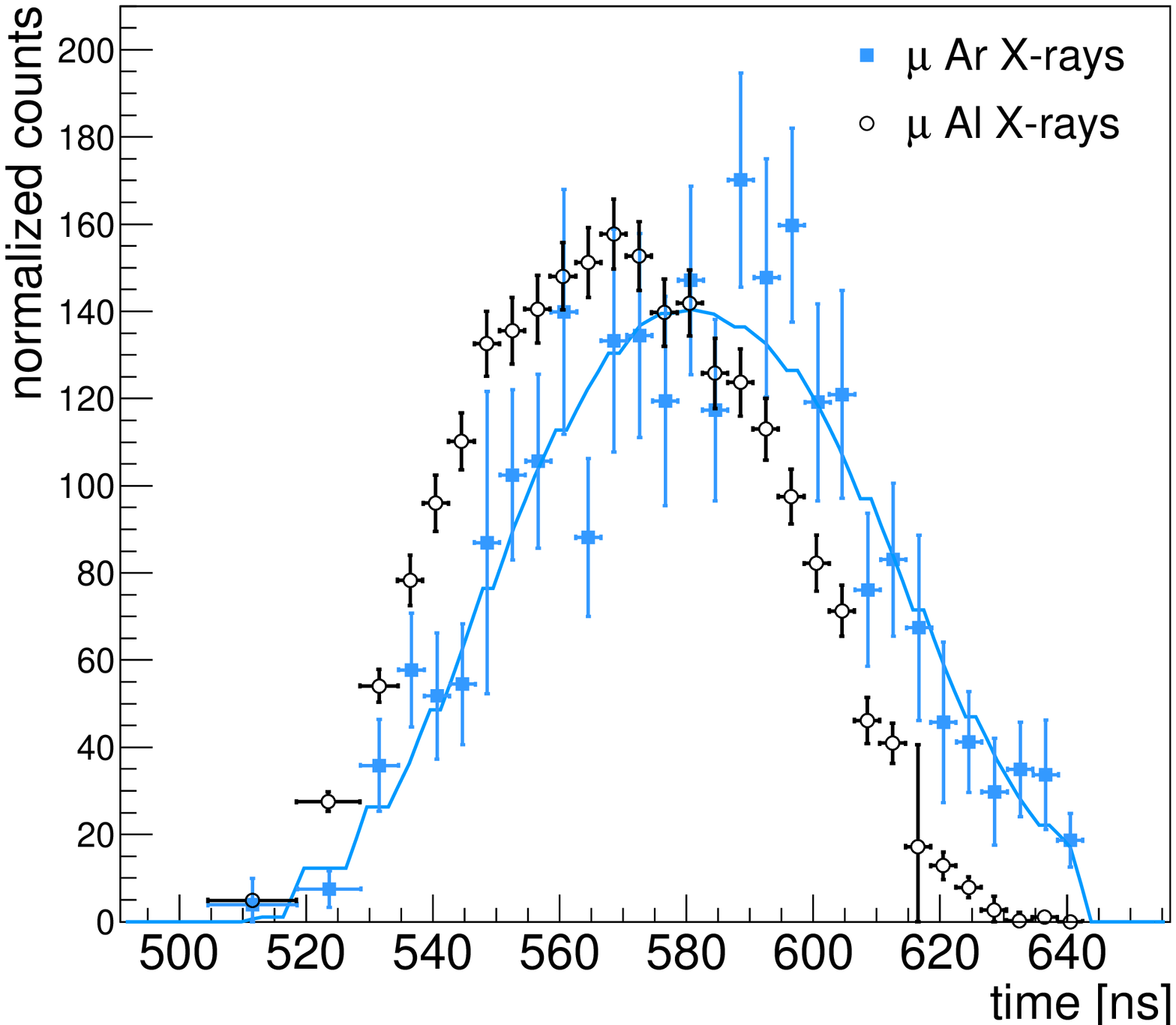}
\caption{Left panel: time evolution of X-rays emission from aluminium and oxygen. Right panel: time evolution of X-rays emission from aluminium and argon. Lines are a fit to the oxygen and argon distributions, see text.}
\label{fig:tevol}       
\end{figure}
Figure \ref{fig:tevol} shows the time distribution of X-rays
originating from muonic aluminium atoms formed in the vessel (346~keV
line) and the X-rays coming from oxygen (133~keV line) and argon (644~keV line). It can be noticed that the aluminium X-ray distribution precedes in
time the contaminant X-ray gas distribution. The data for the argon
sample were acquired with a smaller statistic, resulting in larger fluctuations. The solid line in figures is a fit to the data, as explained later in 
this section. The hypothesis that the X-ray time distributions from
muonic aluminium and oxygen (argon) are 
compatible was rejected performing the Kolmogorov test on unbinned data 
which give zero probability both for oxygen and argon.

The time difference between X-ray distributions from muonic aluminium
and the muonic atoms of the gas admixed to hydrogen has also been studied using a
GEANT4 simulation of the experimental set-up. The simulation
reproduces the structure of the target, the detectors, and space and
time distributions of the muon beam. However, the process of muon
transfer has not yet been implemented and as a consequence only the
timing of the ``prompt'' X-ray emission can be evaluated.
\begin{figure}[!htb]
\centering
\includegraphics[width=0.45\textwidth]{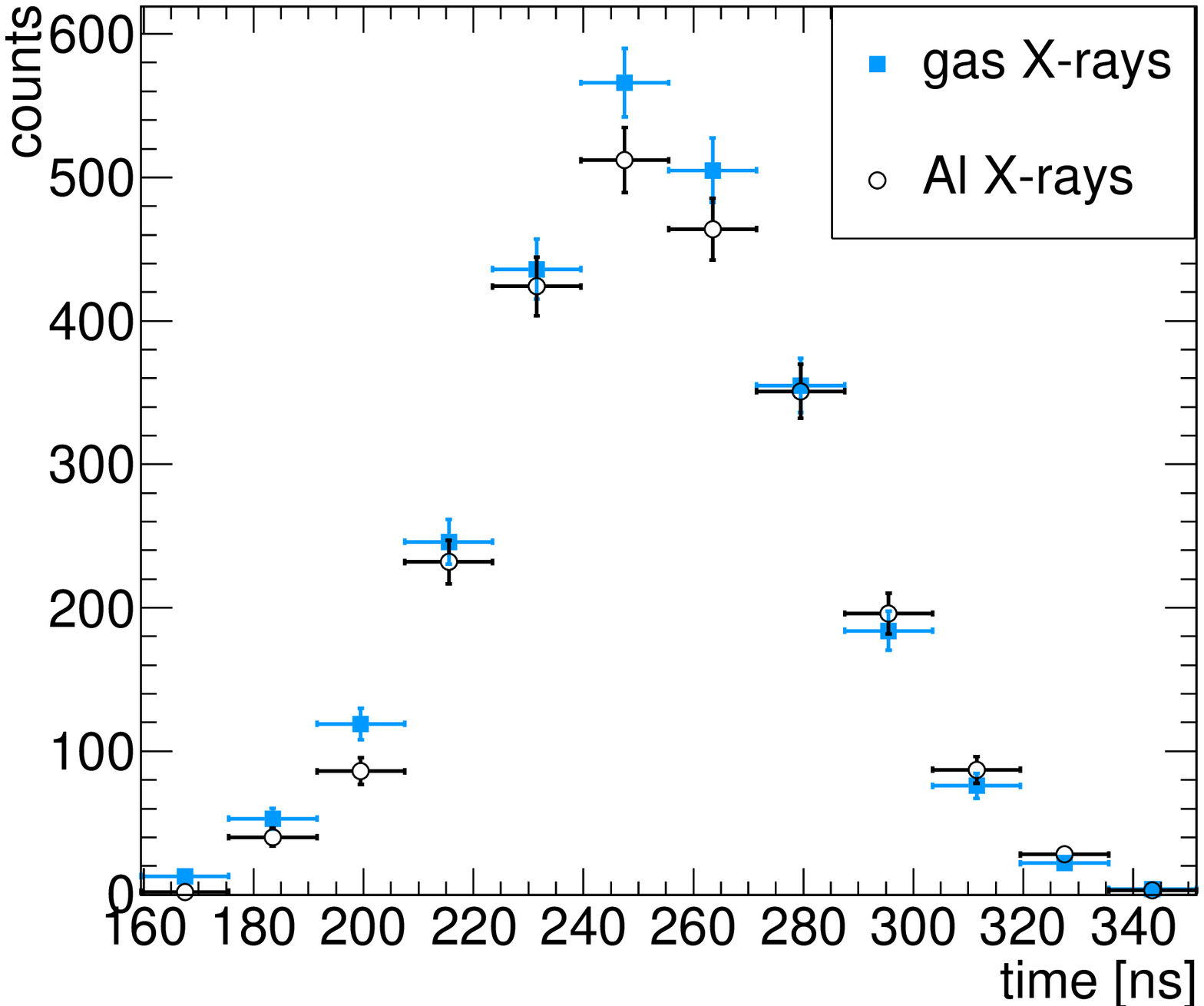}
\includegraphics[width=0.45\textwidth]{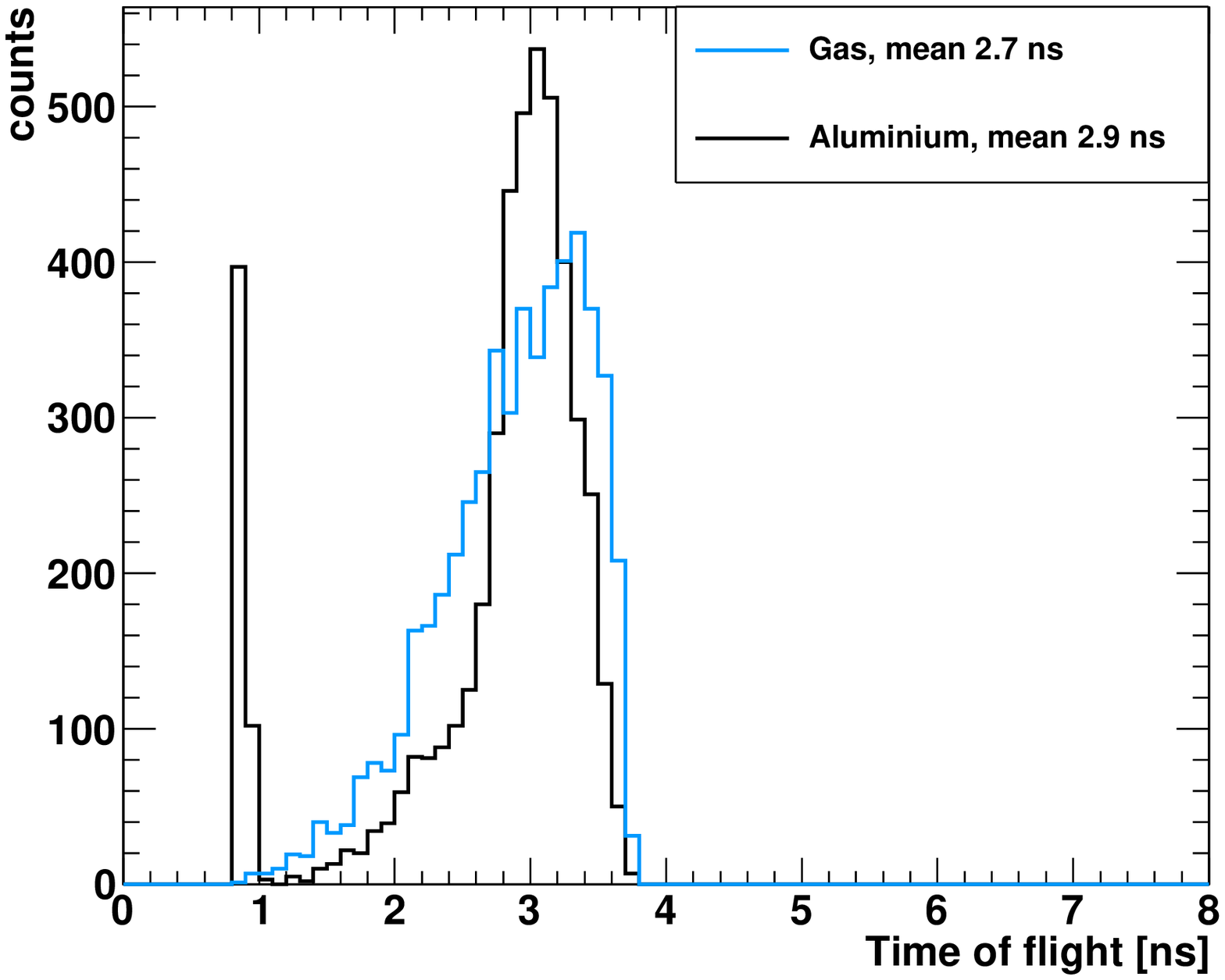}
\caption{Left panel: simulated time evolution of X-rays emission from
  aluminium (full black circles)
  and gas (blue open circles). 
Right panel: time of flight of muons stopping in the
  aluminium (black line) and in the gas (blue line).}
\label{fig:simutevol}       
\end{figure}
Figure \ref{fig:simutevol}, on the left, shows the results of the simulations of the time evolution of muon stopping in the aluminium of
the target and in the gas. The distributions
are almost overlapping; according to a Kolmogorov test, there is a probability of 83\% that the two distribution belong to the same population. The same conclusion can be derived by studying the time of flight
of muons before the stop in the gas or in the aluminium, see
Fig.~\ref{fig:simutevol} on the right, with the mean from the gas distribution at 2.7~ns
and the mean from the aluminium distribution at 2.9~ns. The peak at about 1 ns for aluminium 
corresponds to events stopping in the entrance window. This type of 
events accounts for only about 10\% of the events of muon stop in aluminium. The remaining 90\% of 
muons stops in the lateral walls of the target in a good time 
coincidence of muons stopping in the gas, blue line in figure. 
This effect can be explained by observing that the muon beam momentum was tuned at 61 MeV/c. Hence, most of the muons are able to cross the 3~mm thick aluminium entrance window without stopping. Subsequently, they loose energy in the gas till stopping in hydrogen or eventually reaching the aluminium lateral walls from where, according to the simulation, the most part of the aluminium muonic X-rays are coming. 
 The difference between the distributions obtained from the
 simulation, fig.~\ref{fig:simutevol} on the right, can not account for the time
difference of about 5--10~ns observed in the real data between
aluminium and $CO_2$ (argon) X-rays distributions (see Fig.~\ref{fig:tevol}). This
implies that a physical process like the muon transfer from muonic
hydrogen to  $CO_2$ and argon is responsible for this time difference.

\begin{landscape}
\begin{table}[htb]
  \begin{center} 
    \begin{tabular}{| l | c | c | c | l |}
      \hline
 & Gas mixture & $c_{Z_i} \Lambda_{Z_i}$  terms & Free parameter & Fixed parameters \\
 &              & in eq.~\ref{eq:ldis} &                        &                 \\
      \hline
      Transfer rate to $CO_2$ & $H_2 + 4\% \, CO_2$ & $c_C\Lambda_{pCO_2}$  & $\Lambda_{pCO_2}$ & $c_{C}=(9.5\pm0.3)\times 10^{-4}$  \\
      &              &  $=c_C(\Lambda_{pC}+2\Lambda_{pO})$ & &    $c_O=2c_C$ \\
      & &  & & $c_{d(H_2+CO_2)}=(1.358\pm0.001)\times10^{-4}$\\
      & &  & & $c_{p}=1-c_C-c_O-c_{d(H_2+CO_2)}$\\
      \hline
      Transfer rate to Oxygen & $H_2 + 4\% \,CO_2$ & $c_C\Lambda_{pC} + c_O\Lambda_{pO}$ & $\Lambda_{pO}$ & $\Lambda_{pC}=(5.1\pm 1.2) \times 10^{-10}$s$^{-1}$\\ 
      & &  & & $c_{C}=(9.5\pm0.3)\times 10^{-4} $ \\
      & &  & & $c_O=2c_C$\\
      & &  & & $c_{d(H_2+CO_2)}=(1.358\pm0.001)\times10^{-4}$\\
      & &  & & $c_{p}=1-c_C-c_o-c_{d(H_2+CO_2)}$\\
      \hline
      Transfer rate to Argon &   $H_2 + 2\%  \,Ar$  &  $c_{Ar}\Lambda_{pAr}$ & $\Lambda_{pAr}$ & $c_{Ar}=(5.1\pm0.1)\times 10^{-4}$\\
      & &  & & $c_{d(H_2+Ar)}=(1.355\pm0.001)\times10^{-4}$\\
      & &  & & $c_{p}=1-c_{Ar}-c_{d(H_2+Ar)}$\\
      \hline
    \end{tabular}
        \caption{Parameters used in Eq.~\ref{eq:ldis} for the evaluation of the transfer rate for the different gases.}
  \label{tab:sum}
  \end{center}
  \end{table}
\end{landscape}

Through a study of the time evolution of the X-ray events coming from
the gas, it is possible to determine the average transfer rate in this
particular situation, when in the X-ray spectrum contributions from
the direct muon capture, muon transfer from excited $\mu$p atoms
\cite{thalmann98}, and from the ground-state $\mu$p atoms coexist.
The prompt aluminium signal from the target
material (345 keV line) was chosen as a 
time reference. In the case of $CO_2$ as contaminant in the
gas, Eq.~\ref{eq:ldis} becomes $\lambda_{dis} = \lambda_0 +\phi[c_p\Lambda_{pp\mu} + c_d\Lambda_{pd} + c_O\Lambda_{pO} +c_C\Lambda_{pC} ]$,
where $c_O$ and $c_C$ are the oxygen and carbon
atomic concentrations respectively (with $c_O = 2 c_C$) and $\Lambda_{pO}$, $\Lambda_{pC}$
are the corresponding transfer rates. Since in the condition of this
measurement the transfer rates to oxygen and carbon are unknown, it was decided not only to
evaluate a single average transfer rate to the $CO_2$ molecule ($\Lambda_{pCO_2}=\Lambda_{pC}+2\Lambda_{pO}$) but also to evaluate the transfer rate to oxygen $\Lambda_{pO}$ by using a fixed value of $\Lambda_{pC}$. In this latter case we used
$\Lambda_{pC}=5.1\times 10^{10}$s$^{-1}$ from~\cite{basiladze,dupays05} and the disagreement with
other measurements~\cite{schellenberg} was considered in the
estimation of systematic errors. In the
case of argon a single transfer rate $\Lambda_{pAr}$ was evaluated.
Gas parameters depend on the composition, temperature, and pressure as
described below, and are summarized in Table~\ref{tab:sum};  
values were set as follows: 
\begin{itemize}
\item filling was performed at ``room temperature'' $T\in [288,$ $298]$~K,
 at a pressure $P=(38.00\pm0.25)$ bar, with $CO_2$ concentration of
  $(4.00\pm0.12)$\%, mass weighted, and at room temperature $T\in [288,$ $298]$~K,
  at a pressure $P=(40.00\pm0.25)$ bar, with a concentration of
  $(2.00\pm0.06)$\% mass weighted, for argon;
\item the number densities of the gas mixtures are
  $\phi_{H_2+CO_2} = (4.509\pm0.003)\times10^{-2}$
  and $\phi_{H_2+Ar} = (4.606\pm0.001)\times10^{-2}$
  in LHD atomic units, as derived from
  previous values; 
\item the atomic concentrations of carbon $c_{C} = (9.5\pm0.3)\times 10^{-4}$ and argon 
  $c_{Ar} =(5.1\pm0.1)\times 10^{-4}$ are
  derived from previous values; 
  the deuteron concentrations \\$c_{d(H_2+CO_2)}=(1.358\pm0.001)\times 10^{-4}$  and $c_{d(H_2+Ar)}=(1.355\pm0.001)\times 10^{-4}$ were derived from a laboratory measurement \cite{boschi};
\item remaining data
  were taken from literature and theoretical calculations: $\Lambda_{pp\mu} = 2.01 \times 10^6$ s$^{-1}$~\cite{andreev15}, $\Lambda_{pd} = 1.64 \times 10^{10}$ s$^{-1}$~\cite{adam17}, $\lambda_0 =(4665.01\pm0.14)\times 10^2$~s$^{-1}$~\cite{andreev15,suzuki}.
\end{itemize}

A fit of the $CO_2$ (oxygen and carbon), oxygen, and argon X-rays time evolution can be performed
numerically integrating Eq.~\ref{eq:tevol} by leaving  $\Lambda_{pCO_2}$, $\Lambda_{pO}$, and $\Lambda_{pAr}$ as free
parameters. 
We considered the $K_{\alpha,\beta,\gamma}$  X-ray lines from carbon
and $K_{\alpha,\beta,\gamma}$ from oxygen when evaluating
$\Lambda_{pCO_2}$, the $K_{\alpha,\beta,\gamma}$ from oxygen for
$\Lambda_{pO}$, and $K_\alpha$ from argon for $\Lambda_{pAr}$. 
The fit to the data is shown as solid lines in
Fig.~\ref{fig:tevol} for oxygen and argon. The step-like behaviour is due to the numerical
integration procedure used to calculate the function for each time
bin. The reduced $\chi^2$ of the fit is about 1.6 for oxygen and 0.7 for argon. 
However, for oxygen, it can be noticed that there is a significant 
deviation of the fit with respect to the data around 240~ns that could be
ascribed to unaccounted contribution to the transfer rate from excited 
muonic hydrogen states. 
\begin{table}[htb]
 \begin{center}
    \begin{tabular}{| l | c | c | c |}
      \hline
      [Units: $10^{10} $~s$^{-1}$] & $CO_2$ & oxygen & argon \\
      Transfer rate & $405.3\pm1.5 (stat) ^{+255}_{-111} (sys)$ 
                    & $186.4\pm6.0 (stat) ^{+118}_{-52} (sys)$ 
                    & $289\pm18 (stat) ^{+181}_{-80} (sys)$ \\
      \hline\hline
      \multicolumn{4}{|c|}{Sytematic errors breakdown}\\ 
      \hline
Timing (Sec.\ref{subsec:31})&      $^{+243}_{-81}$ & $^{+112}_{-37} $    & $^{+173}_{-58} $\\ \hline
Background (Sec.\ref{subsec:32}) &      $\pm67 $ &  $\pm31 $        & $\pm48 $ \\ \hline
Method (Sec.\ref{subsec:33}) &          $\pm34 $ &  $\pm16 $         & $\pm24 $ \\ \hline
Temperature (Sec.\ref{subsec:34}) &      $\pm8 $ &  $\pm4 $         & $\pm6 $\\ \hline
Concentration (Sec.\ref{subsec:35}) &    $\pm12 $ &  $\pm6 $         & $\pm9 $\\ \hline
$\Lambda_{pC}$ (Sec.\ref{subsec:36}) &       n/a &  $^{+2.7}_{-3.9} $ & n/a \\ 
\hline
\end{tabular}
    \caption{Result of the transfer rate measurements with systematic errors
      summed in quadrature and breakdown of systematic errors. All
      values are in units 
      $10^{10} $~s$^{-1}$. Systematic errors were evaluated for
      oxygen (third column) and their relative weight rescaled to
      argon and $CO_2$ results (second and forth columns). Details of the systematic errors is given in the indicated sections.}
  \label{tab:errors}
\end{center}
\end{table}

The measured transfer rate to $CO_2$, oxygen and argon are reported in
Table~\ref{tab:errors}. Statistical errors are smaller than estimated
systematic errors (summed in quadrature). A description of systematic
errors evaluation is reported in section~\ref{sec:3}. 
Due to the experimental condition, the mean kinetic energy of the
muonic hydrogen cannot be determined precisely. The whole energy range
 between the mean kinetic energy of thermalized
gas (0.038 eV at 300~K) and the
mean kinetic energy (10 eV) of the sum of the two Maxwellians described previously~\cite{werthmuller98}  should be considered.

\begin{figure}[htb]
  \centering
\includegraphics[width=0.95\textwidth]{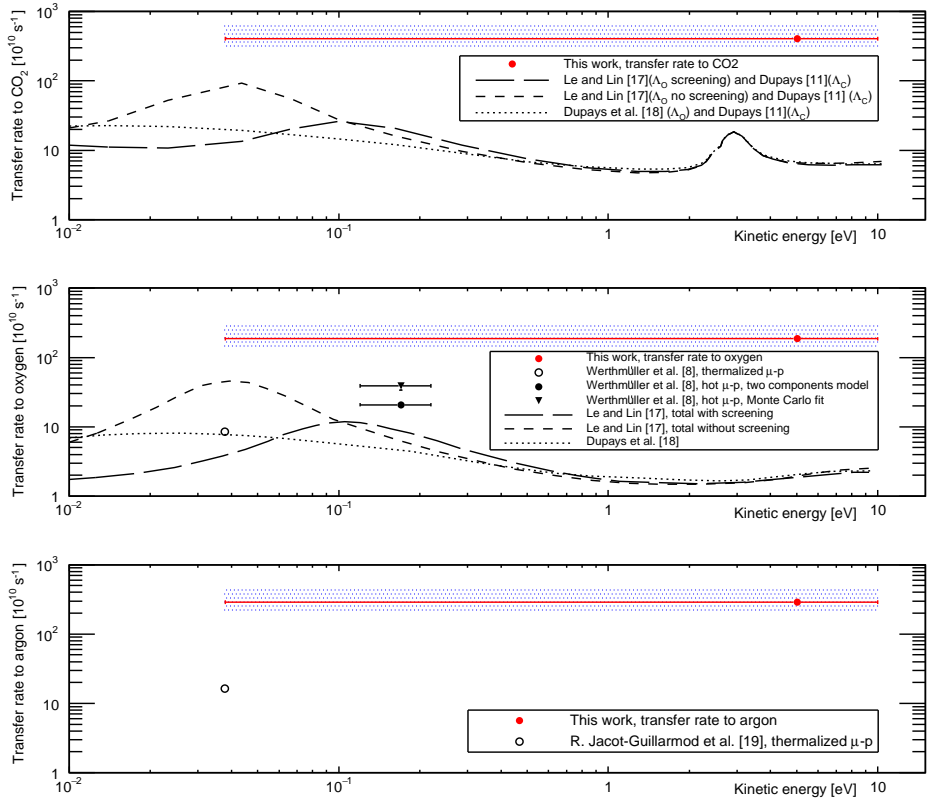}
\caption{From top to bottom: measured transfer rate to
  $CO_2$, oxygen and argon. Shaded regions
  represent the limits of the estimated systematic uncertainties. Horizontal
  lines associated to points represent the energy ranges of the measurements, the points are placed at the arithmetic mean of the interval. Statistical
error bars are included in the points when not visible. Dashed lines
represent theoretical calculations of the transfer rate to
oxygen and to $CO_2$.}
\label{fig:results}       
\end{figure}
Figure \ref{fig:results} shows our results together with previous
measurements and theoretical calculation. In the top panel, our $CO_2$
result is compared to theoretical calculations obtained by combining
transfer rate to oxygen~\cite{lelin,dupays} and to
carbon~\cite{dupays05} in the proportion of 2:1. 
The transfer rate to oxygen estimated by fixing the muon transfer rate
to carbon is shown in the middle panel. 
Results from other measurements obtained at PSI are shown. Open
circles represent measurements made in thermalized conditions at
300~K. Values denoted by black symbols were derived in non-thermalized conditions 
(``hot'' $\mu$p), evaluated using the ``two components'' model and the
``three exponential model'' using a Monte Carlo to determine the best
set of parameters~\cite{werthmuller98}.  The dotted lines show theoretical calculations obtained using a
 computational method to study three-body rearrangement processes~\cite{dupays}.
 Dashed lines represent other theoretical calculations of the transfer rate
 to oxygen obtained by taking or not into account the Thomas-Fermi screening
 in oxygen~\cite{lelin}. 
The bottom panel of Fig.~\ref{fig:results} shows the result obtained for
argon together with a previous measurement~\cite{jacot97}
obtained with thermalized $\mu$p.

 As described previously, our measurement taken with
not-thermalized $\mu$p implies a wide range of possible energy distributions
of which it is not possible to experimentally estimate an average. The full kinetic
energy range has been reported in figures as horizontal ``error
bars''. Where not visible, statistical errors are included in the points. 
Colored bands show the estimated systematic uncertainties described in
the following section.

The value extracted from our measurement of the transfer rates is
larger than previous experiments and of the theoretical calculation
and could be interpreted as an upper limit. Since the current
experimental setup was not devoted to the transfer rate measurement,
these results can be influenced and increased by effects such as the
high level of epithermicity, the presence of a prompt component in the
X-rays distribution and a contribution from muon transfer from excited
states of the $\mu$p atoms. It is worthwhile to notice that the order of
magnitude of the obtained average transfer rates agrees with the
semi-classical calculations for the transfer from the 2S state of $\mu$p
atoms~\cite{fiorentini76}. 

\section{Estimation of systematic uncertainties}
\label{sec:3}
The accuracy of this measurement is limited above all by the
systematic uncertainties, an effect of the conditions under which it
was performed. Estimation of systematic errors was performed using the
muonic oxygen lines. Uncertainties on the transfer rates to $CO_2$ and
argon were extrapolated from this study. 

\subsection{Timing of signals}
\label{subsec:31} 
This measurement of the transfer rate to oxygen is performed by comparing
the time evolution of muonic X-ray from aluminium and oxygen. The
$CO_2$ concentration being very high (4\% by weight), the transfer happens
faster than thermalization and the
time evolution of X-rays spectrum of aluminium and gas are very close and nearly in
coincidence with muon arrival. Consequently, any uncertainty in the
time measurement has a great impact on the measurement. Since both
signals are measured simultaneously with the same setup, there is no
reason to assume an experimental systematics in absolute time
measurement. However, according to the simulation, there is a
difference of 0.2 ns between the X-ray time arrival from the gas
and from the aluminium, with gas X-rays arriving earlier as
explained in previous section. The simulated results can be biased by
the following issue: the muon transfer physical mechanism
is not included in GEANT4 and the beam profile and divergence used in
the simulation have been obtained by RIKEN specialists using
a different setup and positions with respect to the FAMU target.
According to the simulation, the time difference between the mean of
the aluminium and oxygen/carbon X-rays time distributions can vary up to
$\pm$2 ns when taking into account the uncertainties in the beam
profile and divergence, the beam momentum, the position of the target
respect to the beam pipe, the position of the detector with respect to the target.

The effect of such a variation has been tested on the real data by
artificial shifting the average of the aluminium distribution in time
by $\pm$2 ns. The transfer rate obtained fitting the shifted data 
was used to estimate the systematic uncertainty, reported in
Table~\ref{tab:errors}.

\subsection{Background subtraction}
\label{subsec:32}
In order to obtain the number of X-ray events for each time bin, it is
necessary to subtract the background below the X-ray carbon, oxygen, 
and aluminium lines. Background can be determined using spectroscopic
algorithms embedded in ROOT~\cite{ROOT}, parametrizing with
exponential functions the region close to the peaks or by comparing
the energy spectra obtained with pure hydrogen in the target with the
one with $CO_2$ contaminant. The last method was used to obtain the final
results, while the other two were used to study how the result varies
when changing the background estimation method. Moreover, the
normalization of the background introduces small fluctuations in the
total number of observed events that count as systematic effects. The
estimated background subtraction systematic is of the order of 16\%.

\subsection{Systematics of the method}
\label{subsec:33}
Since the transfer time is much smaller than the time interval between
the two beam pulses, the fitting procedure of the time evolution
explained in the previous section can be applied twice for each
trigger. We noticed that, also depending on the background signal
subtraction method, the obtained results differ more than the
statistical fluctuations. Hence, the difference between the results
extracted from the two subsequent muon beam peaks has been treated as
systematic error intrinsic to the method. 
 
\subsection{Temperature}
\label{subsec:34}
The 2014 FAMU target was not equipped with temperature sensors and
the temperature of the gas during filling cannot be precisely
determined. The best estimate of the room temperature is 293$\pm$5
degrees Kelvin. Different temperature of the gas during the
filling procedure implies a different gas density $\phi$. By
propagating the five degrees temperature uncertainty on $\phi$ we
re-fitted the time evolution. A temperature variation of five degrees
corresponds to a variation of the result of
about 2\%.

\subsection{Concentration of gas mixture}
\label{subsec:35}
The gas used were high purity gases, 99.9995 and 99.9999\%
pure, corresponding to a contamination of other gases smaller than 5
and 1 ppm respectively. The simulation proved that such a
contamination implies negligible effects in our results.
However, the gas mixture was prepared by the gas supplier by weight
with a relative error of 3\%. Propagation of this uncertainty in the
fitting formula brings both different atomic concentrations and gas
densities. The overall effect on the final measurement is of about 3\%.

\subsection{Other systematics}
\label{subsec:36}
When measuring the transfer rate to oxygen it is necessary to fix the
value of the transfer rate to carbon, our measurement
was obtained with $\Lambda_{pC} = 5.1 \times
10^{10}$~s$^{-1}$. Since there are conflicting
experimental results \cite{basiladze,schellenberg}, and since the theoretical calculation shows a
variation of the transfer rate to carbon as function of the kinetic
energy, we varied $\Lambda_{pC}$ between $2 \times
10^{10}$ and $10 \times 10^{10}$~s$^{-1}$. This results in  a relative difference of 1.5
and 2\% respectively in the transfer rate. This difference is included the systematics of
the oxygen measurement, Table~\ref{tab:errors}.

Other systematic uncertainties were investigated and were
considered negligible if their effect was smaller
than 1\%. The following items
 were proved to
have negligible effects on the final measurement:
\begin{itemize}
\item literature error in the muon decay time
  (proton-bounded) $\lambda_0$;
\item literature error in the transfer rate to
  $pp\mu$ molecule $\lambda_{pp\mu}$;
\item error in the transfer rate to deuterium
  $\lambda_{pd}$;
\item error in the isotopic composition of hydrogen gas $c_d$;
\item error in the target pressure measurement;
\item effect of the target composition alloy (Al6061) which is a mixture of
  mainly aluminium with a small component of Mg, Si and other heavier
  elements (studied using the simulation).
\end{itemize}
Other consistency checks were also performed on data by selecting
separately $K_\alpha$ and the other $K$ lines and by studying energy
and time spectra obtained using the empty target and a carbon sample.

\section{Conclusions}
\label{sec:4}
The data analysis described here is evidencing the quality of our
experiment's detection system and of the muon beam characteristics
available at the PORT-4 of the RIKEN-RAL facility. 

We have given a
detailed description of the  approach to the  studies of the signal
and of the background noise that have lead to significant results on
muon transfer rate. The first results reported in a previous paper~\cite{adamczak16}
showed a very promising situation for the subsequent steps. In the
present work we have performed a further analysis of those data with
the intent of extracting the transfer rate although to this end the
experimental conditions are far from favorable, demonstrating in this
way the reliability of detection system and of the analysis
method. 

The study has been carried on to test the capabilities of the
target and detectors and to estimate the source of systematic errors
in this type of measurement. Through a careful analysis it was
possible to extract a value for the muon transfer rate from hydrogen
to $CO_2$, oxygen, and argon. The expectations, given the experimental
condition where the transfer arises very quickly well before any
thermalization of the muonic hydrogen, are confirmed. 

The found
transfer rates are higher than what has been previously measured due
to a highly variable level of epithermicity and to the possible
presence of a fraction of prompt component in the muonic X-rays
spectra and where a contribution of the muon transfer from $\mu$p atoms
excited states could also play a role. 

These results further demonstrate the possibility to perform high
precision measurements at the high quality muon beams of the RIKEN-RAL
facility. The next precision measurement of the transfer rate from
muonic hydrogen to nuclei of higher Z  will be performed using a
temperature-stabilized cryogenic hydrogen gas-target with optimized
concentrations of admixed gas. 

\acknowledgments
The research activity presented in this paper has been carried out in the framework of the
FAMU experiment funded by Istituto Nazionale di Fisica Nucleare (INFN). The use of the
low energy muons beam has been allowed by the RIKEN RAL Muon Facility. We thank the RAL
staff (cooling, gas, and radioactive sources
sections) and especially Mr. Chris Goodway, Pressure
and Furnace Section Leader, for their help, suggestions,
professionality and precious collaboration in the set up of the experiment at RIKEN-RAL port 4.

We gratefully recognize the help of T. Schneider, CERN EP division,
 for his help in the optical cutting of the scintillating fibers of the
 hodoscope detector and the linked issues and N. Serra from
 Advansid srl for useful discussions on SiPM problematics.

We thank our colleagues Chiara Boschi and Ilaria Ba\-ne\-schi (IGG, CNR Pisa) for their help in
the measurement of the gas isotopic composition.

A. Adamczak and D. Bakalov acknowledge the support within the bilateral agreement
between the Bulgarian Academy of Sciences and the Polish Academy of Sciences. D. Bakalov,
P. Danev and M. Stroilov acknowledge the support of Grant 08-17 of the Bulgarian Science
Fund.

\end{document}